\documentstyle[preprint,aps,psfig]{revtex}
\begin{document}
\def\be{\begin{equation}}
\def\ee{\end{equation}}
\def\bea{\begin{eqnarray}}
\def\eea{\end{eqnarray}}
\hyphenation{nu-cleo-syn-the-sis ex-pe-ri-ments}

\title{Local Constraints on the Oscillating G Model}
\author{ Jos\'e A. Gonz\'alez, Hernando Quevedo,
 Marcelo Salgado and Daniel Sudarsky\\
\small {\it Instituto de Ciencias Nucleares} \\
\small {\it Universidad Nacional Aut\'onoma de M\'exico} \\
\small {\it Apdo. Postal 70-543 M\'exico 04510 D.F, M\'exico}. }
\maketitle

\abstract{
We analyze the observational constraints  on the
effective Brans-Dicke parameter and on the temporal
variation of the effective gravitational constant  within the context of the
oscillating $G$ model,
a cosmological model based on a massive scalar field non-minimally
coupled to gravity.  We show that these local constraints
cannot be satisfied simultaneously once the values of the free parameters
entering the model become fixed by the global attributes 
of our Universe. In particular, we show that the lower observational bound for
the effective Brans-Dicke parameter and the upper bound of the
variation of the effective gravitational constant lead to a specific value of
the oscillation amplitude which lies well below the value
required to explain  the periodicity of 128 Mpc $h^{-1}$ 
in the galaxy distribution
observed in the  pencil beam surveys.}

\vskip 1cm
 Submission to: {\bf Physical Review D15 (Brief Reports)}

\newpage
\section{Introduction}

The success of the standard cosmological model in describing
the evolution of our Universe, beginning with the era of nucleosynthesis
until the present state, has been confronted with serious
difficulties resulting from the analysis of cosmological data.
At large cosmological scales we find two main problems which 
are not dealt within the framework of the standard (old) cosmological model. 
The first one concerns the cosmological dark matter problem according to which
the luminous matter (baryonic matter and radiation)
content of the Universe represents only a small fraction
of the total matter content. In fact, the inflationary models predicted
that the total energy density $\Omega = 1$, with $\Omega$ given in
terms of the critical energy density \cite{dm3}.
This prediction has recently been given further support
 by observational data resulting from the recent cosmic 
microwave background (CMB) experiments like Boomerang and Maxima and the 
high red-shift supernovae (SNIa) measurements \cite{flat}, 
leading to the conclusion that the average 
energy-density of the Universe is indeed near the critical value. 
Obviously, these observations have increased the importance
of the dark matter problem for the understanding of our Universe.

The second problem is related to the observations that indicate a
periodicity of 128$h^{-1}$ Mpc (where $h$ is the Hubble parameter in 
units of 100 km s$^{-1}$ Mpc$^{-1}$) 
in the galaxy number distribution, observed in deep pencil beams
\cite{Broad,Szalay} in the north and south poles of our galaxy.
This shocking discovery  would, in its simplest interpretation, indicate
 that galaxies in the Universe
are situated on the surface of concentric spheres with the center
situated in our own galaxy. This is in complete contradiction with the basis
of modern cosmology: the cosmological principle of homogeneity and isotropy 
of the Universe. It has been argued \cite{Szalay}, and it seems to be the 
pervading view among researchers in the field, 
that this periodicity could be the result of the appearance 
of an intrinsic length scale in the distribution of matter. However,
we have shown \cite{per} that this explanation is not really  satisfactory as
 there are scenarios of this type that  result in a negligible probability
for such observation
to be obtained in a particular direction.

In a series of works \cite{ssq1,ssq2,ssq3,ssq4} we have investigated an
alternative model based on a massive scalar field which
is non-minimally coupled to gravity. The oscillation of the scalar
field in cosmic time results in a time-dependent
effective gravitational constant. We have shown that this
model leads to predictions which are in good agreement with most of the
observational data. In fact, although this model was originally proposed
\cite{Morik} to explain the observed periodicity in the galaxy number
distribution,  we have shown that it was possible to adjudicate 
most of the energy
density of the Universe to 
the oscillating massive scalar field which, therefore, could be 
regarded as candidate for the non-baryonic nature of the
cosmological dark energy. That is, this model is able to explain
simultaneously both, the problem of the cosmological dark energy and
the problem of the periodicity in the galaxy number distribution.
We have checked that the model satisfies some of the cosmological constraints. 
More precisely, we have seen that the model
reproduces correctly the primordial nucleosynthesis of $^4$He, and is
consistent with the present
value of the energy density of baryonic matter and the age of the Universe.
In this work, we will analyze  the additional constraints following from local
observations, namely, the Viking experiments \cite{viking}, which
impose bounds on the rate of change in time of the effective gravitational
constant and on the effective Brans-Dicke parameter.

In a previous work \cite{ssq5}
we have shown that all but one of the free parameters entering the model are fixed by
the cosmological analysis and that with these values it was not possible
to satisfy the Brans-Dicke bound. In this work, we analyze the possibility
of overcoming this problem by relaxing the single condition freely imposed
in our previous cosmological studies. We will show that even with this
relaxation it is not possible to satisfy the local
constraints  and the  periodicity observations simultaneously. This result
indicates that either the behavior of the scalar field in the presence 
of local inhomogeneities is different from its behavior at large scales 
\cite{ssq5} or that a modified
model would be necessary if we want to explain in a unified way the 
apparent galactic periodicity and the cosmological dark energy.

\section{Constraints on the oscillating $G$ model}

The dynamics of oscillating $G$ model is described by the  Lagrangian:
\be
{\cal L} = \left({ 1\over 16\pi G_0} + \xi \phi^2\right)
\sqrt{-g} R - \sqrt{-g} \left[ {1\over 2}(\nabla \phi)^2
+ V(\phi) \right]\ ,
\label{lag}
\ee
where $G_0$ is Newton's gravitational constant, $\xi$ stands for the
non-minimally coupling constant, $R$ is the scalar curvature, $\phi$ is
the  scalar field  and $V(\phi)$ is a scalar potential which
in its simplest form is taken as the harmonic potential $V=m^2 \phi^2$,
with $m$ the mass of  the scalar field. If we consider
a time-dependent scalar field, the non-minimal
coupling results in a time-dependent effective gravitational constant
$G_{\rm eff} =G_0(1+ 16\pi G_0  \xi \phi^2)^{-1}$. The central feature of the
oscillating $G$ model is that oscillations in the expectation value of $\phi$
induce oscillations in $G_{\rm eff}$ and this leads to oscillations in the
Hubble parameter $H$ which manifest themselves in the redshift measurements
of distant points of the Universe. In turn, the redshift oscillations
give rise to an apparent variation in the density of galaxies.
Consequently, a temporal oscillation of the redshift can be mistakenly
interpreted as a real spatial periodicity in the galaxy number 
distribution. This was
used in previous works \cite{ssq1,ssq2,ssq3,ssq4} to explain the observed 
periodicity of 128$h^{-1}$Mpc in 
the distribution of galaxies in our Universe. To this end, we
analyzed
the Friedman-Robertson-Walker cosmology with a combination of two
non-interacting
perfect fluids (radiation and baryonic matter). From the field equations
we obtain the following expression for the total effective energy density
of the
the system (see \cite{ssq1,ssq3}):
\be
\Omega_{\rm tot} = {1\over 1+ 16\pi\xi\phi_0^2} \left[
 \Omega_{\rm mat} + {4\pi\over 3}\dot{\widetilde\phi}_0^2
+ {4\pi\over 3}\dot{\widetilde\omega}^2\phi_0^2
- 32\pi\xi\phi_0\dot{\widetilde\phi}_0\right] \ ,
\label{omtot}
\ee
where
\be
 \dot{\widetilde\phi}_0 = {d\phi\over d\tilde t}\bigg|_{\rm today} \ , \quad
\tilde t = t H_0\ ,\quad \widetilde\omega = {\omega\over H_0}\ , \quad
\widetilde m^2 = {4\pi\over 3} \widetilde\omega ^2 \ .
\ee
In the above equation, a subscript ``0" stands for the value of the
corresponding
quantity at present time $t=t_0$. The frequency of oscillation $\omega =
m\sqrt{3/4\pi}$
is determined by the period of 128$h^{-1}$Mpc observed in the pencil beam
surveys and
turns out to be $\omega \approx 147 H_0$. Here $\Omega_{\rm matt}= 
\Omega_{\rm bar} + \Omega_{\rm rad}$. Notice that in
Eq.(\ref{omtot}) and for the present analysis we can
neglect the contribution of the photon energy density $\Omega_{\rm rad}$
because
the observations of the cosmic microwave background radiation of 2.725 K
implies that
$\Omega_{\rm rad} \approx 10^{-3}\Omega_{\rm bar}$\cite{footnote}. 
Furthermore, the value of
$\Omega_{\rm bar}$ must lie within the range $[0.01, 0.02]h^{-2}$ determined
by the abundance of the light elements other than $^4$He \cite{copi}. Finally,
for the total energy density we take the value $\Omega_{\rm tot}=1$ in
accordance with
the standard inflationary model and with the recent CMB and SNIa 
observations \cite{flat}.
Consequently, Eq.(\ref{omtot}) can be interpreted as a constraint relating
the initial cosmological values of the scalar field,
$\phi_0$ and $\dot{\widetilde\phi}_0$, and the
coupling parameter $\xi$. Another, in some sense, more realistic approach 
would be to identify $\Omega_{\rm matt}$ with the total amount of clumped 
matter in our Universe which would include besides the baryonic component 
also the so called Cold Dark Matter, leading us to take 
$\Omega_{\rm matt}\sim 0.3$. However, we will see that even this drastic 
change of view  does not alter our conclusions in a significant way.

A further constraint is imposed by the observed redshift-galaxy-count
amplitude
${\cal A}_0\geq {\cal O}(0.5)$, which for the oscillating $G$ model
can be approximated by the expression  \cite{CritStein}
\be
{\cal A}_0 = {16\pi\xi\over \widetilde\omega}\left(
\widetilde\omega^2\phi_0^2 +
\dot{\widetilde\phi}_0^2\right)\ .
\label{amp}
\ee
Here we are considering the additional term
$\dot{\widetilde\phi}_0^2$  which was set to zero in previous analysis
because we want to remove all
the arbitrarily imposed conditions on the model in order to examine whether
all the constraints can be
solved simultaneously. Since the values of ${\cal A}_0$ and
$\widetilde\omega$ are fixed by the pencil beam observations,
Eq.(\ref{amp}) represents a
constraint between the values of $\phi_0$, $\dot{\widetilde\phi}_0$ and $\xi$.

We call Eqs.(\ref{omtot}) and (\ref{amp}) the global constraints of the
oscillating $G$ model because the values of ${\cal A}_0$ and $\widetilde\omega$
are fixed by the  large scale  observations of the galactic
periodicity, and the values of $\Omega_{\rm tot}$ and $\Omega_{\rm mat}$ are
the result of global cosmological observations.

On the other hand, the Solar System local observations impose an upper
bound on the
variation of the  gravitational constant $|\dot G/(GH)|\leq 0.3 h^{-1}$
\cite{CritStein}. For the oscillating $G$ model this yields
\be
\beta = {\dot G_{\rm eff}\over G_{\rm eff} H} = -
{32\pi\xi\phi_0\dot{\widetilde\phi}_0
\over 1 +16\pi\xi\phi_0^2 } \ ,\qquad {\rm with} \qquad  |\beta|\leq 0.3
\label{beta}
\ee
It is well known that scalar-tensor models of the kind defined by the
Lagrangian
(\ref{lag}) can be transformed by means of a conformal transformation into an
effective Brans-Dicke theory. Then, such models can be characterized by an
effective Brans-Dicke parameter $\omega_{\rm BD}^{\rm eff}$ which must satisfy
the lower bound imposed by the Viking experiments \cite{viking},
$\omega_{\rm BD}^{\rm eff}> 3000$. In the case of the oscillating $G$ model we
obtain
\be
\omega_{\rm BD}^{\rm eff} =
{1+16\pi\xi \phi_0^2 \over 128\pi\xi^2 \phi_0^2}\ ,
\label{bd}
\ee
a constraint that relates $\phi_0$ with $\xi$.

Now we proceed to the analysis of the global constraints (\ref{omtot}) and 
(\ref{amp}) and the local constraints (\ref{beta}) and (\ref{bd}). 
In our previous cosmological studies, we were able to satisfy simultaneously the total energy constraint (\ref{omtot}) as well as the nucleosynthesis and 
age constraints, together with the constraints for the amplitude (\ref{amp}) 
and the variation of the effective gravitational constant (\ref{beta}) by 
setting $\dot{\widetilde\phi}_0 = 0$. In fact, in this case the constraint
(\ref{beta}) is 
automatically satisfied ($\beta=0$), whereas the constraints (\ref{omtot})
and (\ref{amp}), together
with the ``plateau hypothesis" \cite{ssq3} that ensures a successful 
nucleosynthesis, fix the values of
the remaining parameters
$\phi_0$ ($\sim 10^{-3}$) and
$\xi$ ($\sim 6)$. The evolution of the model with these 
conditions result in a value for the age of the Universe compatible 
with the standard bounds \cite{age}.

However, as we have shown in \cite{ssq5}, with these values the oscillating
$G$ model
is unable to satisfy the Brans-Dicke limit (\ref{bd}) with
$\omega_{\rm BD}^{\rm eff} > 3000$ (or even the less severe bound 
$\omega_{\rm BD}^{\rm eff} > 500$). The simplest possibility to overcome
this problem
is to relax the condition $\dot{\widetilde\phi}_0=0$ within the range
allowed by the
constraints (\ref{amp}) and (\ref{beta}).
 To this end, we replace the values of
$\phi_0$,  $\dot{\widetilde\phi}_0$ and $\xi$
following from the constraints
(\ref{amp}), (\ref{beta}) and (\ref{bd}) into the total energy constraint
(\ref{omtot}).
Then we obtain
\be
\label{ncon}
f(\omega_{\rm BD}^{\rm eff},\beta,{\cal A}_0  )= 1-\Omega_{\rm mat}
-\frac{2b\,\widetilde\omega\,\omega_{\rm BD}^{\rm eff}{\cal A}_0}{3(a+b)}
+ \frac{b+2\sqrt{a\widetilde\omega {\cal A}_0-b\widetilde\omega^2}} {a} = 0 \ ,
\ee
with 
\be
a=\beta^2 + 4\widetilde\omega^2 \ ,\qquad 
b=-\beta^2+2\widetilde\omega{\cal A}_0
+2\sqrt{\widetilde\omega^2({\cal A}_0^2 -\beta^2)-\beta^2\widetilde\omega{\cal A}_0  }\ ,
\ee
a constraint  that, for a specific value of $\Omega_{\rm mat}$,
 determines the amplitude in terms of the effective Brans-Dicke
parameter and the parameter $\beta$ (recall that 
the frequency $\widetilde\omega$ 
has been fixed by the period of oscillation).
Notice that the initial values
$\phi_0$ and $\dot{\widetilde\phi}_0$ do not appear at all in Eq.(\ref{ncon}). 
In order to investigate the constraint (\ref{ncon}) 
in a systematic way we 
have to solve the algebraic equation (\ref{ncon}) as 
${\cal A}_0= {\cal A}_0(\beta)$. Actually this is equivalent to 
solve the differential equation 
$df(\beta,{\cal A}_0(\beta))/d\beta=0$ (i.e., the resulting differential 
equation $d{\cal A}_0/d\beta = F({\cal A}_0,\beta)$) subject 
to the boundary values $(\beta^i, {\cal A}_0^i)$ such that  
$f(\beta^i,{\cal A}_0^i)=0$, for a fixed $\omega_{\rm BD}^{\rm eff}$ and 
$\Omega_{\rm mat}$. For instance, for
$\dot{\widetilde\phi}_0=0$, and $\omega_{\rm BD}^{\rm eff} = 3000$, 
$\Omega_{\rm mat}= 0.0236$
the pair $(\beta^i =0, {\cal A}_0^i\approx 0.022)$ 
satisfies the constraint (\ref{ncon}) as well as the remaining 
conditions (except of course 
the order of magnitude in the bound on ${\cal A}_0$). 
The result of this
calculation is plotted in Figure 1 for two different values of
$\Omega_{\rm mat}$ within the
range allowed by observations. We see that the range of values 
$(\beta,{\cal A}_0)$ that satisfy $f(\beta,{\cal A}_0)=0$ is extremely 
narrow and that all of the values for the amplitude within this range are
situated well below the lower bound ${\cal A}_0 \geq {\cal O}(0.5)$ imposed
by the redshift-galaxy-count observations. The conclusion is that 
the Brans-Dicke local 
constraint is not compatible with the observed value for the oscillation 
amplitude. 
Further numerical analysis of the constraint (\ref{ncon}) show that an
increase of
the matter density $\Omega_{\rm mat}$ or of the effective Brans-Dicke
parameter leads to even lower values for the amplitude.

We conclude that the relaxation of the original condition
$\dot{\widetilde\phi}_0=0$
does not allow the oscillating $G$ model to satisfy simultaneously both
global and local constraints, and that we have to look for further
generalizations
of this model if we want to consider it as a candidate to explain
the apparent galactic periodicity simultaneously with the nature of the
non-baryonic dark matter
content in the Universe. Needless is to say that had the model 
succeeded in these tests, then it would 
be necessary to confront the oscillating $G$ model to 
further tests in light of the recent CMB and SNIa observations.

Finally, it is worthwhile to emphasize that the generalized view on the 
problem of the galactic periodicity is that perhaps there is no 
problem at all and that such a ``periodicity'' is only 
the result of an excess of power at some characteristic length 
scales. While this could be the case, the simplest analysis on this 
matter shows that the existence of a characteristic distance in the large 
scale distribution is not enough to explain such observations 
\cite{per}, and therefore serious doubts 
arise in taking such a comfortable position. Clearly, 
the observation of galactic periodicity or lack thereoff in directions 
other than those corresponding to the north and south galactic poles will put an 
end to the controversy. On the other hand, 
if the existence of such periodicity in a large number of directions were to 
be confirmed we would be in the uncomfortable situation of having no model 
to account for it, and we would need to resort to variations on the 
oscillating $G$ model presented here as the only type of scenario 
capable of explaining such observations within the context of the cosmological 
principle.
 
\vskip .3cm
{\bf Acknowledgments}

\bigskip
We acknowledge partial support from DGAPA-UNAM Project No.
IN121298 and from CONACyT Projects 32551-E and 32272-E.

\vfill\eject

\begin{figure}
\hspace{3in}
\psfig{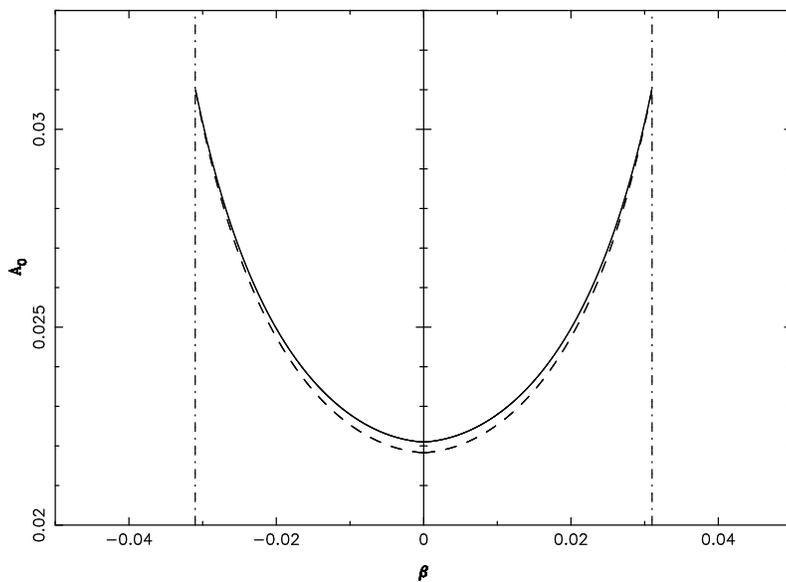}
\caption{The redshift-count-oscillation amplitude ${\cal A}_0$ as a function
of the parameter $\beta$ satisfying the constraint (\ref{ncon}) for 
$\omega_{\rm BD} = 3000$. The solid line corresponds to a value of 
$\Omega_{\rm mat}=0.02366 $  and the dashed line to $\Omega_{\rm mat}=0.04733$.
The dash-dotted lines show the limits for which the 
constraint (\ref{ncon}) is valid. Here we took $H_0= 65\, 
{\rm km\,s^{-1}\,Mpc^{-1}}$.}
\end{figure}


\begin{thebibliography}{99}
\bibitem{footnote}
Obviously we cannot neglect the photon contribution 
when we evolve the equations backwards in time in order to recover the 
nucleosynthesis predictions.
\end{thebibliography}
\end{document}